# PERFORMANCE ANALYSIS AND SPECIAL ISSUES OF BROADBAND STRATEGIES IN THE COMPUTER COMMUNICATION

[1] Dr.S.S.Riaz Ahamed    [2] Dr.D.Mahesh

[1]Principal, Sathak Institute of Technology,  Ramanathapuram,India.

[2.] Asst Professor, Dept of Computer Science & Engg, Mount Zion College of Engineering, Pudukkottai, India.

Email:ssriaz@ieee.org, ssriaz@yahoo.com

## ABSTRACT

Broadband communications consists of the technologies and equipment required to deliver packet-based digital voice, video, and data services to end users. Broadband affords end users high-speed, always-on access to the Internet while affording service providers the ability to offer value-added services to increase revenues. Due to the growth of the Internet, there has been tremendous buildout of high-speed, inter-city communications links that connect population centers and Internet service providers (ISPs) points of presence (PoPs) around the world. This build out of the backbone infrastructure or core network has occurred primarily via optical transport technology. Broadband access technologies are being deployed to address the bandwidth bottleneck for the "last mile," the connection of homes and small businesses to this infrastructure. One important aspect of broadband access to the home is that it allows people to telecommute effectively by providing a similar environment as when they are physically present in their office: simultaneous telephone and computer access, high-speed Internet and intranet access for e-mail, file sharing, and access to corporate servers.

## 1. INTRODUCTION

Once people obtain broadband access to the home, they find that this access needs to be shared with other members of the family using multiple PCs. This includes workers who use laptop PCs at their workplace and desire to be able to use the same laptop at home. As a result, people are installing local-area networks (LANs) in their home. Once this LAN is in place, people want to use it to share files, printers, and devices such as scanners. Once broadband access and home networking reaches critical mass in terms of market penetration, there will be a new class of end-user devices that will enable many new Internet-enabled applications. Already, people are able to perform functions remotely via the Internet: monitoring and controlling their homes, viewing their children who are in day-care centers, checking on live traffic conditions, and playing stereo-quality music over Internet radios. The key drivers for broadband growth, along with the resulting impacts, are summarized in Figure 1. The vision of the broadband home is that broadband multimedia-i.e., video, audio, voice, and data-will be delivered *to and within* the home to personal endpoint devices. Services will be affordable, easy to use, and available to the average family and will be delivered quickly, securely and reliably. Moving forward, all things will be connected [1][5][8][11]-[16].

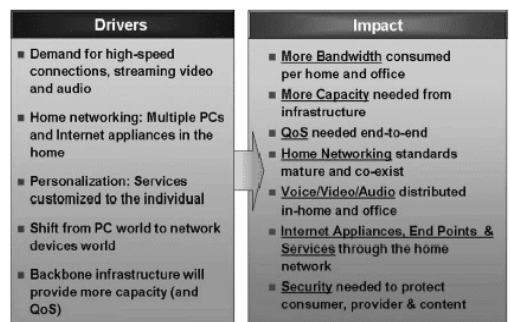

*Figure 1*

Figure 2 shows how broadband connectivity is extended from the core infrastructure to end users' devices such as PCs, personal digital assistants (PDAs), telephones, television sets, and digital cameras. Infrastructure gateway equipment provides broadband access to the packet-based infrastructure. Customer premises equipment (CPE) access gateways extend broadband access connectivity to end-user devices via one or more home networking technologies.

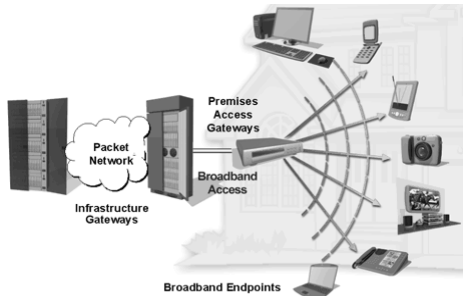

*Figure 2: Broadband Connectivity*

As broadband access becomes available to home users, it is truly changing the way people work and play. Users tired of waiting for Web pages to download using dial-up modem Internet access are signing up for broadband access services. What they find is that not only does broadband access result in fast Web surfing due to higher connection speeds, but it also has several other benefits. The higher connection speeds enable multimedia applications such as real-time Internet audio streaming, posting and displaying digital photographs for friends and family, viewing video clips of news events and movie trailers, and taking virtual tours of hotels and resort areas before making reservations. Because broadband access is always on, unlike dial-up access, there is no wait to connect to the Internet. Thus, people with broadband access tend to leave their personal computers (PCs) turned on and use the Internet for more mundane tasks such as checking television listings and looking up phone numbers-tasks that were not worth the bother when a slow dial-up connection first had to be established. The presence of broadband access also means that the telephone line is no longer tied up when accessing the Internet. This saves the need to purchase a second phone line and enables the user to talk to someone on the phone while accessing information on the Web [5]-[12].

## 2. BROADBAND BUILDING BLOCKS

The challenge of the semiconductor manufacturer is to design silicon and software with a complete solution focus and not just chips or chipsets. As shown in Figure 3, customers have many needs that require a multitude of core competencies for semiconductor manufacturers to satisfy. First of all, customers desire flexible solutions that can accommodate a range of densities to meet requirements scaling from endpoint devices to CPE to carrier-class equipment. Flexibility dictates the need for programmable architecture that can facilitate quick and easy software upgrades to address new features, interoperability issues, performance, and evolving standards.

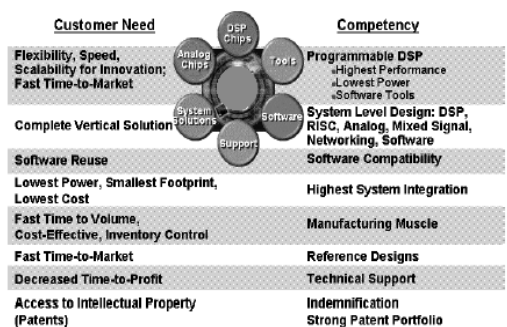

*Figure 3*

For broadband technologies, high-performance digital signal processors (DSPs) are required to implement the various signal-processing algorithms necessary to perform functions such as modulation, voice compression, and video processing. The industry has moved beyond general-purpose DSPs to specialized silicon/software solutions optimized for a particular vertical application. These solutions include the following:

- High-performance analog components, including data converter and amplifiers

- Power management solutions, including power supply control and battery management

- Digital modem technologies for broadband communications

- Radio frequency (RF) wireless technologies for wireless devices

- DSP cores

- RISC processor cores

- Networking interfaces, including Ethernet, Utopia, PCI, USB, 1394, etc.

- Software with well-defined application programming interfaces (APIs), including signal-processing algorithms, protocol stacks, device drivers, real-time operating system pre-ports, network management, and application software

- Voice over packet (VoP) technologies, including voice compression, echo cancellation, tone processing, dial modem, Group 3 facsimile, and telephony signaling

- Networking technologies, including routing, switching, filtering, encryption ,and quality of service (QoS)

These solutions must be very low power to allow battery- or line-powered operation for endpoint devices or to scale in an infrastructure environment where equipment is limited by power consumption and heat dissipation. Cost is always a concern. Customers want semiconductor providers to provide them with the lowest-cost solution, and this is for the entire customer's solution, not just the semiconductor manufacturer's portion. Thus, the semiconductor manufacturer must understand the total build of materials (BOM) cost of the equipment and work at reducing the total BOM. This includes integrating more functionality into the silicon solution and eliminating the need for "glue" logic. It also includes reducing manufacturing costs by making the solution easy to build by minimizing the number of printed-circuit board layers, making the chip package easy to mount and signals easy to route on the printed-circuit board. There is a constant need for cost reduction for mass-market deployment. Customers want to know that the semiconductor's roadmap will offer significant cost reduction for subsequent customer refreshes of the product. The key to facilitating continued cost reduction is to have high-volume manufacturing facilities with leading-edge process technologies coupled with strong system-integration capabilities. To speed customer time to market, semiconductor manufacturers must offer the customers hardware reference platforms that are integrated with software, fully system tested for conformance to industry standards, interoperable with other vendors, and product hardened under real-world conditions. The following subsections look at how building blocks are put together to deliver broadband solutions for infrastructure equipment, premises access gateway equipment, and broadband endpoint devices.

**Infrastructure Equipment**

As shown in Figure 4, broadband infrastructure gateway equipment is responsible for interconnecting broadband access services to the optical core network infrastructure. For multiservices gateways, multicore DSP platforms facilitate the ability to support multiple broadband access technologies as well as traditional voice-grade services. Communications processors containing high-speed processing engines and networking interfaces perform protocol processing and network-management functions. High-speed aggregation logic is required for performing packet processing while providing QoS functions.

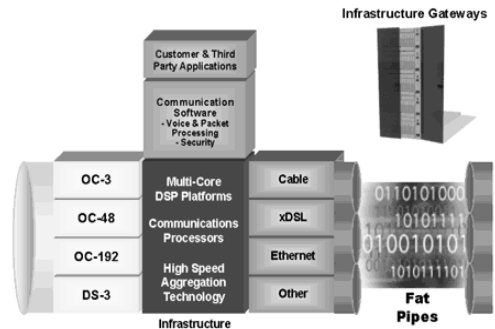

*Figure 4*

Infrastructure gateway equipment providing broadband access is driven by the need to support a large number of end-user connections in a concentrated area (CO or RT unit in the neighborhood) while being constrained on the total power (heat) dissipation.

The concept of solution density has been developed to help service providers and OEMs more clearly understand the technical requirements for implementing high-density products. From a system-engineering perspective, a solution must be evaluated on how the combination of system elements delivers a complete solution with the lowest power and smallest area without compromising quality and features. Solution density refers to the optimization of the overall system architecture, taking into account the following critical elements:

- Power of the solution expressed in milliwatts (mW) per end-user channel
- Density of the solution expressed in end-user channels per square inch
- Cost of the solution, including silicon, hardware, software, and any intellectual property license fees
- System partitioning, including packet aggregation and routing
- Software features that define the functionality of the product
- Network-management capabilities to address high availability and accountability

To engineer an optimal solution, cost, power, and area must be evaluated on a total system basis and must be a function of the features and capabilities supported. For example, the designer must consider the need for external logic, e.g., external memory, aggregation logic, layout/routing issues, etc. In many systems, power (heat dissipation) is the key driving factor especially for high-density solutions. That is, most solutions run out of power in the rack before they run out of board area. Proper functional partitioning is also essential to avoid processing and/or bandwidth bottlenecks in the overall system when scaling to support very large numbers of end-user ports.

*Premises Access Gateway Equipment*

As shown in Figure 5, premises access gateway equipment is responsible for terminating a broadband access pipe from a service provider and making that pipe available to the home or home-office network. Communications processors containing high-speed processing engines and networking interfaces perform protocol processing such as bridging, routing, packet filtering, and firewall operation. Typically, a premises access gateway provides connection to a single broadband access medium-e.g., cable, DSL, or fixed broadband wireless-but may support multiple LAN interfaces such as wired Ethernet, wireless Ethernet, and Bluetooth. VoP technologies are required for derived voice services.

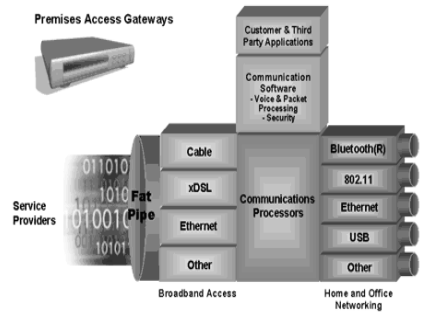

*Figure5*

*Broadband Endpoints*

As shown in Figure 6, broadband endpoint devices come in many forms, such as PDAs, digital cameras, MP3 players, digital television, and IP phones. DSPs perform multimedia processing such as MP3 audio, MPEG-4, and JPEG imaging. High-quality analog components are essential for performing analog- to-digital (A/D) and digital-to- analog (A/D) processing. These consumer devices must be extremely low cost. Devices that are portable handheld devices must be very low power to ensure long battery life. Also, devices that are line powered, e.g., from the infrastructure, Ethernet, or universal serial bus (USB) interface, must adhere to the power constraints of that interface. Thus, low-power devices coupled with power-management technologies are essential.

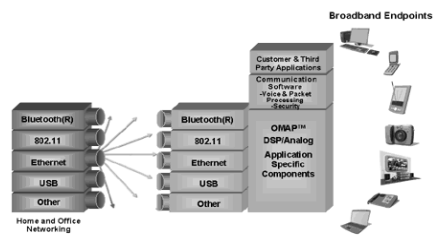

*Figure 6*

Broadband offers users high-speed, always-on Internet access, while offering service providers increased revenue from new, value-added services. Broadband solutions residing within today's broadband communications equipment are complex and require semiconductor manufacturers to integrate a wide variety of innovative technologies to offer low-power, cost-effective system solutions that address the needs of OEMs, service providers, and end users[6][13][15]-21].

## 3. ACCESS TECHNOLOGIES

There are many competing broadband access technologies being brought to bear to address last-mile connectivity, including the following:

- Cable modem
- Digital subscriber line (DSL)
- Fiber
- Cellular wireless
- Wireless Ethernet

*Cable*

As an alternative to existing copper phone wires, cable companies have been providing broadband access by upgrading their cable plant to carry data and voice services in addition to traditional video services. A cable-modem termination system (CMTS) communicates with cable modems located at the customer premises to provide broadband access services. The cable modem typically provides an Ethernet interface to a PC or to a small router when multiple PCs are connected. Today's cable networks generally deliver data with download speeds roughly between 500 kbps and 2 Mbps and upstream speeds of 128 kbps. Newer-generation cable-modem technologies will significantly increase the available bandwidth to further enable interactive applications such as videoconferencing and high-end on-line video.

Internet protocol (IP) telephony is one of the services that can be delivered over coaxial cable. For the cable operators, IP telephony enables them to offer voice services that, to date, have been the domain of the telephone companies.

## DSL

DSL technology is a copper-loop transmission technology for transmitting high-speed data over ordinary telephone wires. A DSL modem is installed at the customer premises and at the central office (CO). Different variants of DSL exist to address different technology trade-offs that can be made regarding different network environments and applications. One of the key trade-offs is distance (referred to as reach) from the CO and data rate. Asymmetrical DSL, or ADSL, is primarily used for residential services. ADSL takes advantage of the fact that there is more crosstalk interference at the CO end of a copper pair than at the subscriber end due to the large bundles of cabling entering the CO. ADSL can provide data rates up to 8 Mbps from the network to the subscriber direction, and up to 1 Mbps from the subscriber to the network direction. The asymmetry of ADSL works well for today's home applications where the majority of bandwidth is consumed in the network to user direction.

Symmetrical DSL, or SDSL, is a cost-effective solution for small and medium enterprises, offering a competitive alternative to T1 and E1 lines. The International Telecommunication Union-Telecommunications Standardization Sector (ITU-T) standard G.991.2, also known as G.shdsl, is a replacement standard for proprietary SDSL. G.shdsl offers data rates from 192 kbps to 2.3 Mbps while providing a 30% longer reach than SDSL.

Very-high-data-rate DSL, or VDSL, can support symmetrical or asymmetrical services. Asymmetrical VDSL is capable of providing data rates to the user of up to 52 Mbps, making it suitable for transporting high-speed applications such as real-time video streaming. The trade-off for this high speed is restricted reach. This requires that the customer be located close to the CO or that the infrastructure access gateway resides outside the CO (and closer to the customers) in a remote terminal (RT).

## Fiber

For new infrastructure buildout, where copper wires are not currently present, the installation of fiber is being employed. Fiber-optic technology, through local access network architectures such as fiber-to-the-home/building (FTTH/B), fiber-to-the-cabinet (FTTCab), and fiber-to-the-curb (FTTC) offers a mechanism to enable sufficient network bandwidth for the delivery of new services and applications. A fiber-optic cable is run from the CO to the neighborhood. Passive optical splitters are used to provide point-to-multipoint connectivity. This is referred to as a passive optical network or PON. In the case of FTTCab or FTTC architectures, the signal is converted to provide connectivity to the subscribers via copper-pair wires. Since these cabinets are collocated in a neighborhood, the copper-pair run is typically less than 3,000 feet; thus enabling high-performance xDSL access to be achieved.

## Cellular Wireless

Next-generation cellular is providing high-speed data capabilities in addition to traditional voice. 2G cellular services only offer data service rates on the order of 9.6 kbps. The emerging 2.5G services will boost available bandwidth to the user and facilitate always-on data services. For 2.5G networks, there are two primary technologies: general packet radio service (GPRS) and enhanced data rates for GSM and TDMA (IS-136) evolution (EDGE). Third-generation (3G) wireless communication technologies support even higher data rates. The packet switching is IP-based, making for efficient routing of data from the Internet through the carrier's gateway. The higher bandwidth should allow for better integration of voice, data, and video signals. Delivery of data services over cellular offers the promise of ubiquitous high-speed data access, including while in moving vehicles.

*Wireless Ethernet*

In addition to cellular-based wireless data services, wireless Ethernet, traditionally a home and enterprise networking technology, is being used for broadband access in public areas such as airports, hotels, sports arenas, convention centers, and coffee shops. This allows users to take their laptop and PDA devices with them and to use a common access technology to deliver high-speed Internet services in their office, home, and while on the road.

*Home and Enterprise Networking Technologies*

While most corporations today have some form of wired Ethernet LANs to address their networking needs, most homes do not have any form of networking infrastructure. There are several competing home networking technologies, including the following:

- Ethernet
- HomePNA
- HomePlug
- Bluetooth®
- Wireless Ethernet

Ethernet is the most ubiquitous LAN technology and as such, very low-cost Ethernet adapters exist for PCs and other devices. However, installing Ethernet cabling in existing homes is expensive as it involves labor-intensive work to snake cables through existing walls, install outlets, and repair drywall. As such, installation of Ethernet cabling is typically relegated to new construction. As an alternative, technology has been developed to use existing phone wiring to run LAN traffic simultaneously with voice. The Home Phone Networking Alliance (HomePNA) defines standards for interoperability using this technology. Unfortunately, most homes have a limited number telephone jacks for access to the wires. Thus, the expense of adding new wires must still be tackled. Technology has been developed to use existing home AC wiring to run LAN traffic. As most rooms have multiple AC outlets, there is access to the LAN from practically anywhere. This still requires the device accessing the LAN to be tethered, as it must plug into the AC outlet.

As an alternate to wired networks, wireless standards exist, including Bluetooth and wireless Ethernet (wireless LAN or WLAN). Bluetooth was developed to replace the need for interconnect cabling between devices for short-range and relatively lower data rates. Wireless Ethernet is a standard developed by the Institute of Electrical and Electronics Engineers (IEEE) (802.11) that preserves Ethernet compatibility and data rates. It is gaining wide traction for home, enterprise, and public access networking.

The current standard for wireless Ethernet is 802.11b, and it offers 11 Mbps transmission rates using direct sequence spread spectrum (DSSS) technology. The standard, also known as Wi-Fi™, is widely used in offices, campuses, and homes. Radio transmission is in the 2.4-GHz band. The 802.11a variant of the standard operates in the 5-GHz frequency band and offers transmission rates up to 54 Mbps using orthogonal frequency division multiplexing (OFDM) technology in which the devices determine a set of noninterfering frequencies, multiplex these frequencies, and use them in parallel to achieve greater bandwidth. A recent addition to the 802.11 standard is 802.11g, which extends DSSS operation to 22 Mbps and also supports OFDM operation in the 2.4-GHz frequency band.

Wireless standards must address potential transmission interference with other devices, including microwaves, cordless telephones, and other wireless standards that operate at the same frequency. Also, since it is wireless, solid encryption is required for security purposes.

| Networking Standard | Description | Type | Installation Requirements | Maximum Data Rate Mbps |
|---|---|---|---|---|
| 802.3 Ethernet | Highest capacity; commodity hardware | Wired | New wires | 10/100 |
| 802.11 Ethernet | Wireless Ethernet | Wireless | No wiring required | 11/22/54 |
| HPNA | Home networking using existing telephone wires | Wired | Some new wires | 1/10/32 |
| Bluetooth | Short range cable replacement | Wireless | No wiring required | 1 |
| HomePlug | Home networking using existing AC power | AC Wires | No new wires | 10 |

*Table 1*

Broadband communications consists of the technologies and equipment required to deliver packet-based digital voice, video, and data services to end users. Broadband offers users high-speed, always-on Internet access, while offering service providers increased revenue from new, value-added services. Broadband solutions residing within today's broadband communications equipment are complex and require semiconductor manufacturers to integrate a wide variety of innovative technologies to offer low-power, cost-effective system solutions that address the needs of OEMs, service providers, and end users[1][17][19][21]-[26].

## 4. MEDIA SERVICES

Broadband media services is the seamless, customized, "on demand" creation and delivery of multimedia services to homes, businesses, and mobile users, including entertainment services (movies, interactive games, broadcast TV), infotainment (e-learning, online training) through high-speed Internet protocol (IP) networks. "Broadband media" is sometimes called "streaming media" because the services, or "content," that is delivered via broadband networks is digitized, and received by users of the content in continuous real-time "streams." Broadband content is digitized and accessed utilizing IP, the standard protocol used for Internet access today. In fact, high-speed IP access through digital subscriber lines (DSL) that utilize existing voice lines for high-speed transmissions, is the foundation of the broadband media services network, and DSL is available in many parts of the world today. DSL is a group of increasingly high-speed technologies that enables fast Internet access in homes and businesses. DSL "always on" connections will also form the basis of the sophisticated broadband media services networks of tomorrow. Fast Internet access barely scratches the surface of the powers of broadband, DSL, and IP technology, which, combined in broadband media services, will connect people and businesses around the world like never before. Broadband media services will put the consumer in total control by enabling personal, custom, on-demand viewing of entertainment, e-learning, video games, and other types of content. Individuals will choose what they want to hear, see, or be entertained by on their own, and people will no longer have to plan around preconceived broadcast schedules for home entertainment. Eventually, we will decide our own schedules for much of our entertainment.

The Next-generation network, the first truly data-oriented broadband network supporting broadband media services, will be all IP, meaning all access to the network will occur via IP standards. The evolution of the broadband media services network can be characterized by six different transitions:

- Transition from a dial-up-like circuit-switched network to a data-oriented network
- Transition from connectivity to service-creation platforms
- Transition from a copper-based network towards an all-optical network
- Convergence of fixed networks
- Convergence of mobile and fixed networks
- Transition to IP version 6 (Ipv6) networks

In short, next-generation networks will evolve to better reflect the requirements of broadband media services. In practice this means bringing IP and other associated network functionalities in the network closer to the customers. The DSL technology and network components that enable high-speed IP access and basic broadband media services exist today, and will remain the foundation of the next-generation broadband media services network:

The major components of a broadband IP access network and next generation broadband media services network are

- high-speed DSL access multiplexers (DSLAM) equipment, located in the operator central 0ffice (CO) and/or in remote locations close to end-users
- broadband access servers
- DSL modems in the home and/or office providing fixed local-area networks (LAN) and wireless LAN (WLAN) network access
- Network- and service-management and provisioning products
- loop management for managing DSL services in the local telecom loop
- IP network security and authentication products for network security and user identification

**Components**

The broadband media services components can have varied functionality with just a minor change in the presentation of the feature, which is required for a modular and scalable solution as new services are created and consumer demand for additional services evolves. Essentially, broadband media services allows consumers to customize their viewing via network control devices. Each set of devices or "boxes" can support a unique content lineup map, which enables consumers to select and pay for only the media that interest them.

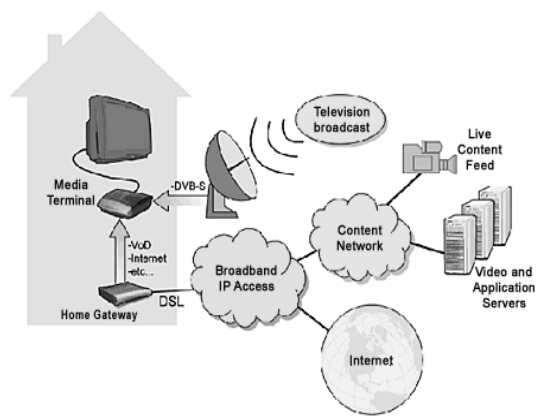

Figure 7. Broadband Media Services for the home

**Video Encoders**

Video encoders are devices that create digital video. Input to the encoders can be analogue video or a Digital Video Broadcasting Group (DVB) multiplex. Both are required because some video content will be statically loaded from video tapes and some content will be captured from a satellite (DVB) multiplex. Video encoders that are used to deliver broadband media services most often allow for the creation of MPEG content and have the ability to support IP multicast at varying bit rates, as well as the ability to decrypt video streams to remove conditional access.

**Video Servers**

Video servers perform two major functions. First, they act as content repositories for the material being streamed. Second, they are responsible for streaming out video and audio using the desired format and network protocol. Video servers can be scaled from streaming 20 to over 5,000 simultaneous video streams. Video servers generally support several different transport protocols for video delivery.

**Interactive Television Application**

Interactive TV (ITV) applications consist of many different applications. The core of the system is the application framework and the data-handling capabilities of the back-end systems. Highly scalable for add-on features, the fundamental applications in an ITV system are customer relationship management (CRM) software modules that track customer usage, profiles, buying characteristics, and application subscription information and create billing events that-can be exported to various billing systems. Applications that typically run on the application framework are VOD, time-shifted TV, web access integrated with video applications, e-mail, personalized user interfaces, broadcast multichannel TV, and pay-per-view applications. Variants of these fundamental applications include channel blocking; parental controls; instant web access associated with viewing preferences for an enhanced, interactive viewing experience; video special offers; and targeted advertising.

**Set Top Box and Customer Premises Equipment**

The set top box and customer-premises equipment (CPE) are devices that are placed in consumer homes or offices, either as two separate devices or as one device combining the home or office gateway functionality required for broadband media services delivery to fixed and wireless devices. A set top box is an electronic device that serves as an interface between a television set and a broadband network, providing VOD and interactive multimedia services. CPE is any type of network device that sits in the home or office of the consumer, as opposed to the central network office or remote sites. User connections to broadband media services are made through modems and media terminals in the home and office, while the main infrastructure lies in the back-end networks, invisible to the end-user[2][4][8]-[16].

**5. BROADBAND OVER DSL**

**Digital Subscriber Line**

Digital Subscriber Line (DSL) is a commonly used term these days. Almost every single home in the urban US had a telephone line. It was almost a given. There is an almost seamless transition from Dial-Up internet for those users who use an ISP (Internet Service Provider), like America Online (AOL) or Netzero experience to DSL which guarantees much higher speeds.

DSL uses the non-voice frequencies of the telephone line to transmit data. So, natural voice conversations are not affected by DSL, unlike the annoying Dial-Up, where if your grandma picked up the phone while you were connected to the internet, she would be greeted by the deafening tones over a hoarse telephone line, or with a roar from you, because your internet line got disconnected!

### Cable Internet

But there was a silent giant lurking out there, riding over one of the other most popular components of every home, with great degree of transparency for home users. This was Cable TV. People receiving Cable TV programs over their televisions, could now plug in a splitter, much like they would, to share the cable connection between two TVs or illegally between neighbors. Instead of having another TV on an end of the splitter, they would plug in a black box, called the Cable Modem.

Cable Modem had a strangle hold over the US Markets back then and to an extent even now! Somehow, it just seemed to be a natural synergy between the cable TV and the internet. Of course, every home had a Cable TV connection. Though, the telephone is even more popular than Cable TV in homes, Cable has dominated the US Markets. For a premium, if you could obtain high speed internet connection, that would be the path of least resistance for most consumers. The only requirement would be a spot close to your TV to put the modem, so that those thick black wires do not have to travel through the home like rattlesnakes.

## DSL Concept

DSL uses the non-voice frequencies of the telephone line to transmit data. Getting a DSL connection most often requires you to enter into a commitment for at least a year, if you wish to get good rates; I mean the money in this case. Services are also partitioned by the maximum achievable data rates. So, if you wish to get higher speeds, you have to pay more. There are several kinds of DSL services which are offered. At the time of this writing, the ADSL family is in vogue and we are tending towards the VDSL market. Before this piece gets to become a cluster of abbreviations, I will stall it right here. ~S **A** ~T stands for Asymmetric and V stands for Very High Speed. DSL, as you may remember is, Digital Subscriber Line.

Asymmetric DSL is based on the premise that while surfing the internet, you get a lot more data coming into your browser than what you send to it. Even if you are typing a huge post to a blog, what you type is considered as an offline activity, since everything you type is stored locally until you hit the send or post button. The optimal approach for someone would be to type everything into a local editor and then cut-paste onto the 'Blog This' window. I have already mentioned that in the days of dial-up internet, especially in places where it is really tough to get the connection over noisy lines, it would be to your advantage to do as much homework before connecting on to the internet. With DSL, which is a dedicated, **" *always ON* "** connection, this is not so much of a problem. The data rate in the receiving side is about 10-30 times the data rates you can get while sending.

## DSL Components

DSL technology consists of two ends, one side is what most of us who have DSL in our homes are familiar with, the Modem or Customer Premise Equipment (CPE). The other side, which end users do not have to worry about, is the Exchange or the Central Office (CO) side.

The distance between the CPE and the CO is called *loop length* . This is a crucial determinant of what rates you can get at your home which is inversely proportional to the distance from the Central office of your DSL service provider, such as Verizon, SBC or Bellsouth. There are various challenges along the way, including but not limited to channel noise, echo, crosstalk, interference from Amateur Radio and other impairments, that can reduce the net data rate to your home.

On the bright side, DSL is not so much impacted by other users in your area, except that there maybe some degradation due to Crosstalk, if there are several DSL connections within the same binder of telephone cables. This is however, not as much as what Cable Modem users might see during 'peak hours', when several people are logged on at the same time. But it is as of today, a fact that cable modem users are seeing consistently higher data rates at off peak hours than DSL for the same price, here in the US, which is why Cable is still the most popular broadband medium, however, that has shown signs of changing in the near future.

SBC Yahoo is already providing DSL at very competitive rates of about 15 bucks a month for a 1 year term. Add this to a basic local phone service and you pay less than 40 bucks a month for atleast one year. This compares much more favorably to Cable costs which exceed 70 bucks even for Basic Cable plus High speed Internet.

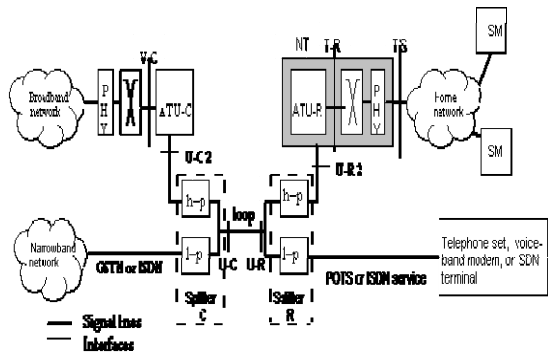

Figure 8 : DSL Block Diagram [Source: ITU 992.1 Standards]

ATU-C = Central Office (CO)
ATU-R = Customer Premise Equipment (CPE)
h-p = High Pass that allows data transmission
l-p = Low Pass that allows voice conversations
loop = Channel or line in between CPE and CO
Narrowband Network: For phone conversations or Dial Up networking services
Broadband Network: For broadband services such as DSL or Cable & data related services like VoIP.

## ADSL Challenges and Opportunities

ADSL technology is evolving rapidly with newer standards and other performance optimizations. The concept of Multi-Carrier Modulation is a standard in ADSL. Think of this as the band of non-voice frequencies being broken down into several bins, each of which can carry a certain number of bits. The information to be sent is broken down between these bins. There are some bins that can carry more information than the other depending on the channel conditions. The aggregate of the bits carried in the bins every second is a measure of the data rate in Kilobits per second. Performance is improved by increasing the number of bins which can carry meaningful data. This led to an evolution of ADSL2 and ADSL2+ standards. The latter innovation has now lead to data rates as high as 30Mbps in the receive direction, also called as *'downstream'* from the Central Office. The *'upstream'* rates to the Central Office are as high as 1.2Mbps. So there is a great push to keep extending the range of Downstream data rates, since there is more and more demand for Video and Audio over DSL. This means more downstream information. It is likely when operators deploy ADSL2+ at competitive rates, DSL will gain much more popularity. In emerging markets like in China , Taiwan and India , majority of the work is in ADSL1, which has matured considerably in terms of standards specifications[4][11][13][15]-[22].

## 6. WiMAX

WiMax (Worldwide Interoperability for Microwave Access) is a wireless broadband technology, which supports point to multi-point (PMP) broadband wireless access. WiMax is basically a new shorthand term for IEEE Standard 802.16, which was designed to support the European standards. 802.16's predecessors (like 802.11a) were not very accommodative of the European standards, per se. The IEEE wireless standard has a range of up to 30 miles, and can deliver broadband at around 75 megabits per second. This is theoretically, 20 times faster than a commercially available wireless broadband. The 802.16, WiMax standard was published in March 2002 and provided updated information on the Metropolitan Area Network (MAN) technology. The extension given in the March publication, extended the line-of-sight fixed wireless MAN standard, focused solely on a spectrum from 10 GHz to 60+ GHz.This extension provides for non-line of sight access in low frequency bands like 2 - 11 GHz. These bands are sometimes unlicensed. This also boosts the maximum distance from 31 to 50 miles and supports PMP (point to multipoint) and mesh technologies. The IEEE approved the 802.16 standards in June 2004, and three working groups were formed to evaluate and rate the standards. WiMax can be used for wireless networking like the popular WiFi. WiMax, a second-generation protocol, allows higher data rates over longer distances, efficient use of bandwidth, and avoids interference almost to a minimum. WiMax can be termed partially a successor to the Wi-Fi protocol, which is measured in feet, and works, over shorter distances.

Fixed wireless is the base concept for the metropolitan area networking (MAN), given in the 802.16 standard. In fixed wireless, a backbone of base stations is connected to a public network. Each of these base stations supports many fixed subscriber stations, either public WiFi hot spots or fire walled enterprise networks.

These base stations use the media access control (MAC) layer, and allocate uplink and downlink bandwidth to subscribers as per their individual needs. This is basically on a real-time need basis.

The subscriber stations might also be mounted on rooftops of the users. The MAC layer is a common interface that makes the networks interoperable. In the future, one can look forward to 802.11 hotspots, hosted by 802.16 MANs. These would serve as wireless local area networks (LANs) and would serve the end users directly too. WiMax supporters are focusing on the broadband ~Slast mile~T in unwired areas, and on back-haul for WiFi hotspots. WiMax is expected to support mobile wireless technology too, wireless transmissions directly to mobile end users. WiMax changes the last mile problem for broadband in the same way as WiFi has changed the last one hundred feet of networking. WiMAX has a range of up to 31 miles, which can be used to provide both campus-level network connectivity and a wireless last-mile approach that can bring high-speed networking and Internet service directly to customers. This is especially useful in those areas that were not served by cable or DSL or in areas where the local telephone company may need a long time to deploy broadband service.

## WiMax Architecture

WiMax offers a rich feature set and flexibility, which also increases the complexity of service deployment and provisioning for fixed and mobile networks.

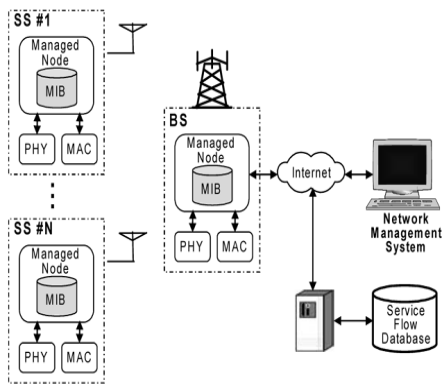

*Figure 9  WiMax Management Information Base [Source:**Intel.Com**]*

The figure 9 shows the management reference model for BWA (Broadband Wireless Access ) networks. This consists of a Network Management System (NMS), some nodes, and a database. BS and SS managed nodes collect and store the managed objects in an 802.16 MIB format. Managed objects are made available to NMS' using the Simple Network Management Protocol (SNMP). When a customer subscribes to the WiMax service, the service provider asks the customer for the service flow information. This would include number of UL / DL connections with the data rates and QoS parameters. The customer also needs to tell the kind of applications that he proposes to run. The service provider then proceeds to pre-provision the services and enters the information in the Service Flow Database.

## WiMax Standards, Technology and Challenges

WiMax, like DSL/cable is standards based and enables vendors to interoperate with one another. The IEEE standard 802.16 was specified for Worldwide Interoperability for Microwave Access (WiMax). This standard has been further revised for 2-11GHz fixed (802.16.a-2004) and 2-6GHz portable (802.16e) wireless solutions. The WiMax Forum has been chartered with taking the standards (IEEE/ETSI) and implementing interoperable solutions for WiMax. The forum is comprised of a group of diverse companies, ranging from silicon vendors to equipment manufacturers to wireless carriers, all having a stake in the deployment of WiMax. The WiMax Forum conducts a number of interoperability events, to bring vendors on common ground. There will also be a WiMax certification for interoperability with multiple vendors, similar to the TR-067 Interoperability tests for DSL or WiFi certification. Typical WiMax equipment would contain a baseband-PHY processor and the MAC network processor besides memory and other peripherals. WiMax uses Orthogonal Frequency Division Multiplexing (OFDM) for modulation in its physical layer, which bundles data over narrowband carriers transmitted in parallel at different frequencies. The same technique has been used as Discrete Multi-tone Modulation (DMT) in ADSL. OFDM makes WiMax scalable for a fluctuating user base, since the spectrum can be dynamically reallocated (range: 1.25-20 MHz) with variations in the number of subscribers. In addition, OFDM improves resilience to interference and outdoor environment, and improves the signal to noise ratio at the terminals[3][11]-[19][45]-[49].

There are a number of challenges facing WiMax including:

- Concerns with WiMax base station market growth due to bias towards voice networks
- RF interference and attenuation
- Government regulations, spectral licensing/usage management

**Future of WiMax**

Like any new technology, WiMax faces challenges. But once these are overcome, this technology, especially the portable WiMax, has very exciting prospects of unwiring the high speed internet access technologies. Although it will not entirely replace incumbent broadband technologies, it will serve as a viable complement to them.

## 7. CONCLUSION

Broadband access is not only about providing the pipes to carry the traffic on the Internet but also about how the traffic will be carried. As was said at the outset, the Internet has changed our sense of time itself. The increasing importance of the Internet has brought about dramatic changes in the way goods and services are produced and distributed to end users. The Internet continues to play a significant role in shaping the new economy by enabling firms to communicate and conduct business on a global basis without regard for location or asset size. Investment in high technology continues to serve as an engine of strong productivity growth for the U.S. economy. Federal Reserve Board Chairman Alan Greenspan expects this trend to continue in the years ahead. The increasing penetration of broadband access among both business and consumer users significantly augments this trend. There are many players in the same space, and the winners will be the technologies and the companies that clearly define their products and services, know and satisfy their customers, are forward-looking and flexible, and set the pace for the industry. Technological innovation and, in particular, the spread of IT has revolutionized the conduct of business during the past 10 years and has resulted in large increases in productivity. The surge over the past few years in business capital spending is a direct result of the higher rates of return brought about by the application of new technologies. The pace of innovation may have temporarily slowed down, but it is expected to resume soon as companies begin to exploit the largely untapped potential for e-commerce, especially in the B2B sector, from which much of the growth is expected to come. The demand for high-speed bandwidth continues to grow at a fast pace, driven mostly by growth in data volumes, as the Internet and related networks become more central to business operations. Today's telecom industry is undergoing a bandwidth shortage driven mostly by the continuing explosion of the Internet and data markets. The rapid growth of distributed business applications; the proliferation of private networks, e-commerce, and bandwidth-intensive applications (such as multimedia, videoconferencing, and VOD); and the continuing deregulation and privatization of the telecommunications networks throughout the world all help fuel the demand for bandwidth. Moreover, an increasing number of teleworkers are fueling the demand for second and third lines for fax and Internet dial-up. To meet this explosive demand for bandwidth and to capitalize on this growing data opportunity, many data CLECs are targeting small business, SOHOs, and teleworkers in the selected areas of the country in which they are operating.

In only a few short years, all Internet appliances or electronic devices will be able to access the Internet and obtain the same content as PCs. The continued evolution of Internet appliances will continue the trend of making regulatory boundaries and lines between customer equipment and network services and communications and broadcasting very murky. It will become increasingly difficult to classify particular services, industries, and providers in nice little boxes, where it will be easy to tell if a service is subject to regulation or not. With accessibility to technology no longer an issue, how and when content and content providers will change to accommodate this ubiquitous access from any type of Internet-enabled device has yet to be answered.

## 8. REFERENCES

1) James Gaskin, Broadband Bible (Bible), ISBN13: 9780764569517, ISBN10: 0764569511, Powell's Publications, 2003, Pp 43-195.

2) Yassini Rouzbeh, Planet Broadband, Publisher: Cisco Pres, ISBN: 1587200902, 2004, Pp 110-210.

3) Jeffrey G. Andrews,Fundamentals of WiMAX: Understanding Broadband Wireless Networking, Published in 2007,Prentice Hall, ISBN 0132225522 , Pp 87-152.

4) Agis, E., Mitchel, H., Ovadia, S., Aissi, S., Bakshi, S., Prakash, I., Kibria, M., Rogers, C., and Tsai, J. 2005. "Global Interoperable Broadband Wireless Networks: Extending WiMax Technology to Mobility". Intel Technology Journal.

5) Intel. 2004. "Broadband Wireless: The New Era in Communications". INTEL White Paper.

6) Kurose, J. and Ross, K. 2005. *Computer Networking: A Top-Down Approach Featuring the Internet*. Addison Wesley: New Jersy.

7) Odinma, A.C. 2006. "Next Generation Networks: Whence, Where and Thence". *Pacific Journal of Science and Technology*. 7(1):10-16.

8) A. Bakre and B.R. Badrinath. I-TCP: Indirect TCP for mobile hosts. Proceedings of the 15th International Conference on Distributed Computing Systems (ICDCS). May 1995, Pp 48-90.

9) H. Balakrishna, et al. A comparison of mechanisms for improving TCP performance over wireless links. Proceedings of ACM/IEEE Mobicom, Pp. 77–89, September 1997.

10) H. Balakrishna et al. Improving TCP/IP performance over wireless networks. Proceedings of ACM/IEEE Mobicom. November 1995, Pp 110-134.

11) L. Blunk and J. Vollbrecht. PPP extensible authentication protocol (EAP). IETF RFC 2284. March 1998.

12) C. Borman et al. Robust header compression (ROHC): Framework and four profiles: RTP, UDP, ESP, uncompressed. IETF RFC 3095. July 2001.

13) R. Braden et al. Resource reservation protocol. IETF RFC 2205. September 1997.

14) L. Brakmo and L. Peterson. TCP Vegas: End to end congestion avoidance on a global internet. IEEE Journal on Selected Areas in Communication, Pp 1465–1480, October 1995.

15) H. Gossain et al. Multicast: Wired to wireless. IEEE Communications Magazine, 40(6):116–123, June 2002.

16) M. Gudmundson. Analysis of handover algorithm. Proceedings of the IEEE Vehicular Technology Conference, May 1991.

17) M. Gudmundson. A correlation model for shadow fading in mobile radio. Electronics Letters, Pp 14–21, November 1991.

18) J. Hodges and R. Morgan. Lightweight directory access protocol (v3): Technical specifications. IETF RFC 3377, September 2002.

19) K. Xu et al. TCP—Jersey for wireless IP communications. IEEE Journal on Selected Areas in Communications, Pp 747–756, May 2004.

20) J. Zander, and SL. Kim. Radio Resource Management for Wireless Networks. Artech House, 2001,Pp 112-197.

21) K. Leung et al. Mobility management using proxy mobile IPv4. draft-leung-mip4-proxy-mode-01.txt, Internet Draft, June 2006.

22) Broadband Powerline Communications: Network Design by Halid Hrasnica, John Wiley, ISBN:9780470857410 , 2004,Pp 89-176.

23) Integrated Communications Management of Broadband Networks by David Griffin,Crete University Press, Heraklion, ISBN 960 524 006 8, Pp 77-145.

24) G. W. Bernas and D. M. Grieco, "A Comparison of Routing Techniques for Tactical Circuit-Switched Networks," *Proc. ICC '78*, pp. 23.5.1-.5.

25) B. W. Boehm and R. L. Mobley, "Adaptive Routing Techniques for Distributed Communications Systems," *IEEE Trans. on Communications*, vol. COM-17, pp. 340-349 (June 1969).

26) B. Widrow, et al., "Adaptive Antenna Systems," *Proc. IEEE*, vol. 55, pp. 2143-2159 .

27) B. Widrow, et al., "Adaptive Noise Cancelling: Principles and Applications," *Proc. IEEE*, vol. 63, pp. 1692-1716

28) P. J. Smith, M. Shafi, and H. Gao, "Quick Simulation: A Review of Importance Sampling Techniques in Communications Systems," *IEEE J. on Selected Areas in Comm.*, Vol. 15, pp. 597-613 (May 1997).

29) E. S. Sousa, J. Silvester, and T. D. Papavassiliou, "Computer-Aided Modeling of Spread Spectrum Packet Radio Networks," *IEEE J. Sel. Areas in Comm.*, Vol. 9, pp. 48-58 (January 1991).

30) S. Chennakeshu and G. J. Saulnier, "Differential Detection of pi/4 - Shifted-DQPSK for Digital Cellular Radio," *IEEE Trans. on Vehicular Technol.*, Vol. 42, pp. 46-57 (February 1993).

31) B. Cutler, "Effects of Physical Layer Impairments on OFDM Systems," *RF Design Magazine*, May 2002, pp. 36-44.

32) M. Da Silva, "Data Above Voice Systems," *Mobile Radio Technology*, October 2003, pp. 16-27.

33) C. C. Davis et al., "Flexible Optical Wireless Links and Networks," *IEEE Communciations Magazine*, March 2003, pp. 51-57.

34) A. Wegener, "CRF: Subtract Peaks to Add Value," *Wireless Design and Development*, October 2005.

35) D. A. Wiegandt, et al., "High-Performance Carrier Interferometry OFDM WLANs: RF Testing," *Proc. 2003 IEEE ICC*.

36) J. E. Wieselthier, G. D. Nguyen, and A. Ephremides, "Algorithms for Energy-Efficient Multicasting in Static Ad Hoc Wireless Networks," *Journal on Special Topics in Mobile Networking and Applications* (MONET), Vol. 6, pp. 241-263, June 2001.

37) D. Wong and T. J. Lim, "Soft Handoffs in CDMA Mobile Systems," *IEEE Personal Communications*, December 1997, pp. 9-17.

38) IEEE. Standard 802.16-2004. Part16: Air interface for fixed broadband wireless access systems. October 2004.

39) Odinma, A.C., L.I. Oborkhale, and M.M.O. Kah. 2007. "The Trends in Broadband Wireless Networks Technologies". *Pacific Journal of Science and Technology*. 8(1):118-125.

40) Long, C. 2000. *Mobile and Wireless Design Essentials*., Wiley Publishing, Inc.: New York.

41) Odinma, A.C. 2004. "The Convergence of Telecom Networks and Migration Strategies for Operators". NSE, Electr. Div., National Conference. 6 – 7 October 2004.

42) Fong, B., Ansari, N., Fong, A., Hong, G., and Rapajai, C. 2003. "On Scalability of Fixed Broadband Wireless Access Network Deployment". *IEEE Radio Communications*

43) Olexa, R. 2004. "Implementing 802.11, 802.16, and 802.20 Wireless Networks: Planning, Troubleshooting, and Operations". *Communications Engineering*. Newnes: London.

44) Pereira, M. 2000. "Fourth Generation: Now, It Is Personal". *Proceedings of the 11th IEEE Intern. Symp. on Personal, Indoor and Mobile Radio Communications*. London, UK, Sept. 2000.

45) WiMAX Forum. Mobile WiMAX—Part I: A technical overview and performance evaluation. White Paper. March 2006. www.wimaxforum.org.

46) WiMAX Forum. Mobile WiMAX—Part II: A comparative analysis. White Paper. April 2006. www.wimaxforum.org.

47) WiMAX Forum. WiMAX Forum Mobile System Profile. 2006–07.

48) IEEE. Standard 802.16e-2005. Part16: Air interface for fixed and mobile broadband wireless access systems—Amendment for physical and medium access control layers for combined fixed and mobile operation in licensed band. December 2005.

49) IEEE 802.16 — The IEEE 802.16 Working Group on Broadband Wireless Access Standards

50) WiMAXPro.com - The Original WiMAX News, Information & Whitepapers